\documentclass[aip,amsmath,amssymb,reprint]{revtex4-1}

\usepackage{amsmath,amssymb,color,lipsum,multirow}
\usepackage{graphicx}
\usepackage{dcolumn}
\usepackage{bm}
\usepackage{txfonts}
\usepackage{mathtools}
\usepackage{hyperref}
\usepackage{soul}

\begin{document}

\title{Thermodynamic uncertainty relation to assess biological processes}
\author{Yonghyun Song}
\affiliation{Korea Institute for Advanced Study, Seoul 02455, Korea}
\author{Changbong Hyeon}
\email{hyeoncb@kias.re.kr}
\affiliation{Korea Institute for Advanced Study, Seoul 02455, Korea}

\begin{abstract}
We review the trade-offs between speed, fluctuations, and thermodynamic cost involved with biological processes in nonequilibrium states, and discuss how optimal these processes are in light of the universal bound set by the thermodynamic uncertainty relation (TUR). 
The values of the uncertainty product $\mathcal{Q}$ of TUR, which can be used as a measure of the precision of enzymatic processes realized for a given thermodynamic cost, are  suboptimal when the substrate concentration $[S]$ is at the Michaelis constant ($K_\text{M}$), and some of the key biological processes are found to work around this condition. 
We illustrate the utility of $\mathcal{Q}$ in assessing how close the molecular motors and biomass producing machineries are to the TUR bound, and for the cases of biomass production (or biological copying processes) we discuss how their optimality quantified in terms of $\mathcal{Q}$ is balanced with the error rate in the information transfer process.  
We also touch upon the trade-offs in other error-minimizing processes in biology, such as gene regulation and chaperone-assisted protein folding. A spectrum of $\mathcal{Q}$ recapitulating the biological processes surveyed here provides glimpses into how biological systems are evolved to optimize and balance the conflicting functional requirements. 
\end{abstract}
\maketitle

From the perspective of thermodynamics, 
biological machineries required to sustain living matters are at work out of equilibrium. 
Their operations to carry out specific functions, which must defy the effect of noise inherent to cellular environment, necessarily incur thermodynamic costs \cite{AlbertsBook,Bustamante2005PhysicsToday,Mugnai2020RMP,QianBook,KolomeiskyBook,Hyeon11BJ}. 
As such, the quest for the relationships between thermodynamic costs and information processing has been a recurring theme in biological sciences \cite{Hopfield1974,Ninio1975,bennett1982IJTP,sagawa2012PTP,Lan2012NaturePhysics,flamholz2013PNAS,sartori2014PLOSCOMP,Cao2015NatPhys,Zhang2020NatPhys,hong2020JSM,basan2020Nature}.  

Recently, a new class of thermodynamic principle, called the thermodynamic uncertainty relation (TUR), has been derived, giving us quantitative ideas on the tradeoff between the thermodynamic costs and uncertainty (or precision) of dynamic processes generated in nonequilibrium \cite{barato2015PRL,Gingrich2016PRL,horowitz2019NaturePhys,hasegawa2019PRL,hasegawa2019PRE}. 
For continuous time Markov jump processes or overdamped Langevin processes under constant nonequilibrium drives, 
TUR in nonequilibrium steady states (NESS) is written as 
\begin{align}
\mathcal{Q} =\Delta S_{\rm tot}(t)\frac{\langle \delta X(t)^2\rangle}{\langle X(t)\rangle^2}   \geq 2k_B. 
\label{eq:TUR_original}
\end{align}
Here, $k_B$ is Boltzmann's constant, which will be set to $k_B=1$ throughout the paper. 
$\Delta S_{\rm tot}(t)=\langle \Delta s_{\rm tot}(t)\rangle$ is the mean total entropy production from both the system and its environment ($\Delta s_{\rm tot}(t)=\Delta s_{\rm sys}(t)+\Delta s_{\rm env}(t)$), calculated over an ensemble of trajectories generated for time $t$. 
$\Delta S_{\rm tot}(t)$ is deemed the thermodynamic cost (or dissipation) to maintain the dynamical process in nonequilibrium for $t$ \cite{horowitz2019NaturePhys}.
$X(t)$ is a time-integrated current-like observable with odd parity ($X(-t)=-X(t)$) that best captures the \emph{functional} feature of the process with $\langle X(t)\rangle$ and $\langle \delta X(t)^2\rangle(=\langle X^2(t)\rangle -\langle X(t)\rangle^2)$ being its mean and variance. 
For the process generated from an enzyme reaction, 
one could \emph{choose} the net number of catalyses or product formations that have occurred for time interval $t$ ($\Delta n(t)=n(t)-n(0)$) as a natural output observable of the process, $X(t)=\Delta n(t)$. 
Upon time reversal, i.e., $t\rightarrow -t$, $\Delta n(-t)=n(-t)-n(0)=n(0)-n(t)=-\Delta n(t)$, satisfying the odd parity.  
If the enzyme exhibits motility along a filament, the net displacement of the enzyme ($\Delta x(t)=x(t)-x(0)$) can be taken as the output observable. 
From the central limit theorem, the entropy production and the square of relative uncertainty grows with time $t$ as 
$\Delta S_{\rm tot}(t) \propto t$ and $\langle \delta X(t)^2\rangle/\langle X(t)\rangle^2\propto1/t$. 
As a result, the product between the two quantities, termed the uncertainty product $\mathcal{Q}$, is a \emph{time-independent} constant. 
The inequality sign of TUR (Eq.\ref{eq:TUR_original}) constrains the value of $\mathcal{Q}$, specifying the minimal dissipation for a given uncertainty, or the minimal uncertainty for a given amount of dissipation.

Originally, the TUR was conjectured and derived for overdamped, and continuous time Markov jump processes at long time limit ($t\rightarrow \infty$) under time-independent non-equilibrium drives \cite{barato2015PRL,Gingrich2016PRL,shiraishi2016PRL,Pigolotti2017PRL,dechant2020PNAS}. 
The steady-state TUR in finite-time as in Eq.\ref{eq:TUR_original}, which is more relevant for the analysis of real systems, was also proven \cite{horowitz2017PRE,gingrich2017inferring,dechant2018JSM,dechant2018multidimensional,hasegawa2019PRE} and experimentally verified \cite{pietzonka2017PRE}.  
The TUR has recently been extended to more general settings, for the fluctuations in first passage times \cite{gingrich2017PRL}, for underdamped condition \cite{lee2019PRE,van2019PRE}, in the presence of magnetic field \cite{chun2019PRE}, under time-dependent drives \cite{koyuk2018JPA,koyuk2019PRL,koyuk2020PRL}, and for arbitrary initial states \cite{liu2020PRL}. 
Furthermore, generalized versions of the relation with less tight bound have been derived from the large deviation principles \cite{Proesmans:2017} and fluctuation theorems \cite{potts2019PRE,proesmans2019hysteretic,hasegawa2019PRL}.

Since the majority of biological processes can be described by employing the language of Markov jump processes on cyclic kinetic network or overdamped Langevin processes, 
we confine ourselves in this perspective to the version of TUR described in Eq.\ref{eq:TUR_original}.
In this paper, we will first clarify the physical significance of the bound set by the TUR from a perspective of the stochastic  thermodynamics. 
Next, by quantifying the uncertainty product $\mathcal{Q}$ for various biological processes with particular emphasis on biological motors and biomass producing machineries, 
we provide our unified perspective on how the functionally critical features, such as thermodynamic cost, reaction speed, and fluctuations of those processes 
are balanced in light of TUR \cite{Hwang2018JPCL}. 
For the case of biomass production processes that transfer the sequence information of DNA or RNA to its downstream, the error in information transfer is another key quantity to be tightly regulated \cite{Song2020JPCL}. 
We will address how the error probability ($\eta$) is balanced with the dynamical features integrated in the $\mathcal{Q}$ by the error correcting machineries. 
Including other dynamical processes associated with information processing that can also be analyzed to determine the value of $\mathcal{Q}$, we will construct a spectrum of $\mathcal{Q}$ to provide an  idea of how different biological processes comprising cellular activities balance their functional needs under the fundamental constraint dictated by the TUR.  

\section*{Physical significance of Thermodynamic uncertainty relation} 
Before discussing the TUR, we review the basics of the stochastic thermodynamics \cite{seifert2005PRL,Seifert2012RPP,park2018JKPS}, which is closely linked to the TUR \cite{seifert2019stochastic}.   
In the stochastic thermodynamics, the thermodynamic quantities are first defined in a trajectory-based manner. 
When the probability of observing a certain trajectory $\Gamma\equiv \{{\bf x}_t\}=(x_0,x_1,\ldots x_t)$ is given by $P(\Gamma)$ with $\int \mathcal{D}\Gamma P(\Gamma)\equiv \int dx_0dx_1\ldots dx_tP(x_0,x_1,\ldots,x_t)=1$, a trajectory-based observable $\Theta=\Theta(\Gamma)$ has its  average value formally obtained by taking the average over all possible trajectories as $\langle \Theta\rangle=\int \mathcal{D}\Gamma P(\Gamma)\Theta(\Gamma)$. 
The basic premises of the stochastic thermodynamics is to (i) ``define" the stochastic entropy
\begin{align}
s_{\rm sys}(t)\equiv -\log{P(x_t)}. 
\label{eqn:stochasticentropy}
\end{align}
and (ii) the trajectory-based total entropy production is given by the ratio of probabilities associated with the original path of the trajectory and its time-reversal \cite{seifert2005PRL}, such that 
\begin{align}
\Delta s_{\rm tot}(\Gamma)&=\log{\left(\frac{P[\Gamma]}{\tilde{P}[\tilde{\Gamma}]}\right)}=\log{\left(\frac{P[\Gamma|x_0]P(x_0)}{\tilde{P}[\tilde{\Gamma}|{\tilde{x_0}}]P(\tilde{x_0})}\right)}
\label{eqn:FB_ratio}
\end{align}
where $P[\Gamma|x_0]$ is the conditional probability of forward path starting from $x_0$ and evolving along $\Gamma$, and the $\tilde{P}[\tilde{\Gamma}|{\tilde{x_0}}]$ is for its time-reversal with $\tilde{x_\tau}\equiv x_{t-\tau}$ and $\tilde{\Gamma}\equiv (\tilde{x_0},\tilde{x_1},\cdots,\tilde{x_t})=(x_t,x_{t-1},\ldots,x_0)$. 
The expression is further decomposed into the entropy production for the reservoir and the system \cite{seifert2005PRL}, 
\begin{align}
\Delta s_{\rm tot}(\Gamma)=\underbrace{\log{\left(\frac{P[\Gamma|x_0]}{\tilde{P}[\tilde{\Gamma}|{\tilde{x_0}}]}\right)}}_{=\Delta s_{\rm res}(t)}+\underbrace{\log{\left(\frac{P(x_0)}{P(x_t)}\right)}}_{=\Delta s_{\rm sys}(t)}. 
\end{align}
Note that the change of the system entropy ($\Delta s_{\rm sys}(t)$) is the difference between the initial and final points (Eq.\ref{eqn:stochasticentropy}), and that the entropy production from the reservoir is the log-ratio of the conditional probabilities of the forward and time-reversed paths  \cite{Schnakenberg1976RMP,kurchan1998fluctuation}. 
The total entropy production follows from the formal definition of trajectory-based ensemble average, such that 
\begin{align}
\Delta S_{\rm tot}&=\langle \Delta s_{\rm tot}(t)\rangle=-\int \mathcal{D}\Gamma P(\Gamma)\log{\left(\frac{P[\Gamma]}{\tilde{P}[\tilde{\Gamma}]}\right)}\nonumber\\
&=\langle \Delta s_{\rm res}\rangle+\langle \Delta s_{\rm sys}\rangle. 
\end{align}
with 
\begin{align}
\langle &\Delta s_{\rm sys}\rangle=s_{\rm sys}(t)-s_{\rm sys}(0)\nonumber\\
&=-\int dx_tP(x_t)\log{P(x_t)}+\int dx_0P(x_0)\log{P(x_0)}
\end{align}
and 
\begin{align}
\langle \Delta s_{\rm res}(t)\rangle=\int \mathcal{D}\Gamma P[\Gamma|x_0]\log{\left(\frac{P[\Gamma|x_0]}{P[\tilde{\Gamma}|{\tilde{x_0}}]}\right)}. 
\end{align}

\begin{figure}[t]
	\includegraphics[width=0.9 \linewidth]{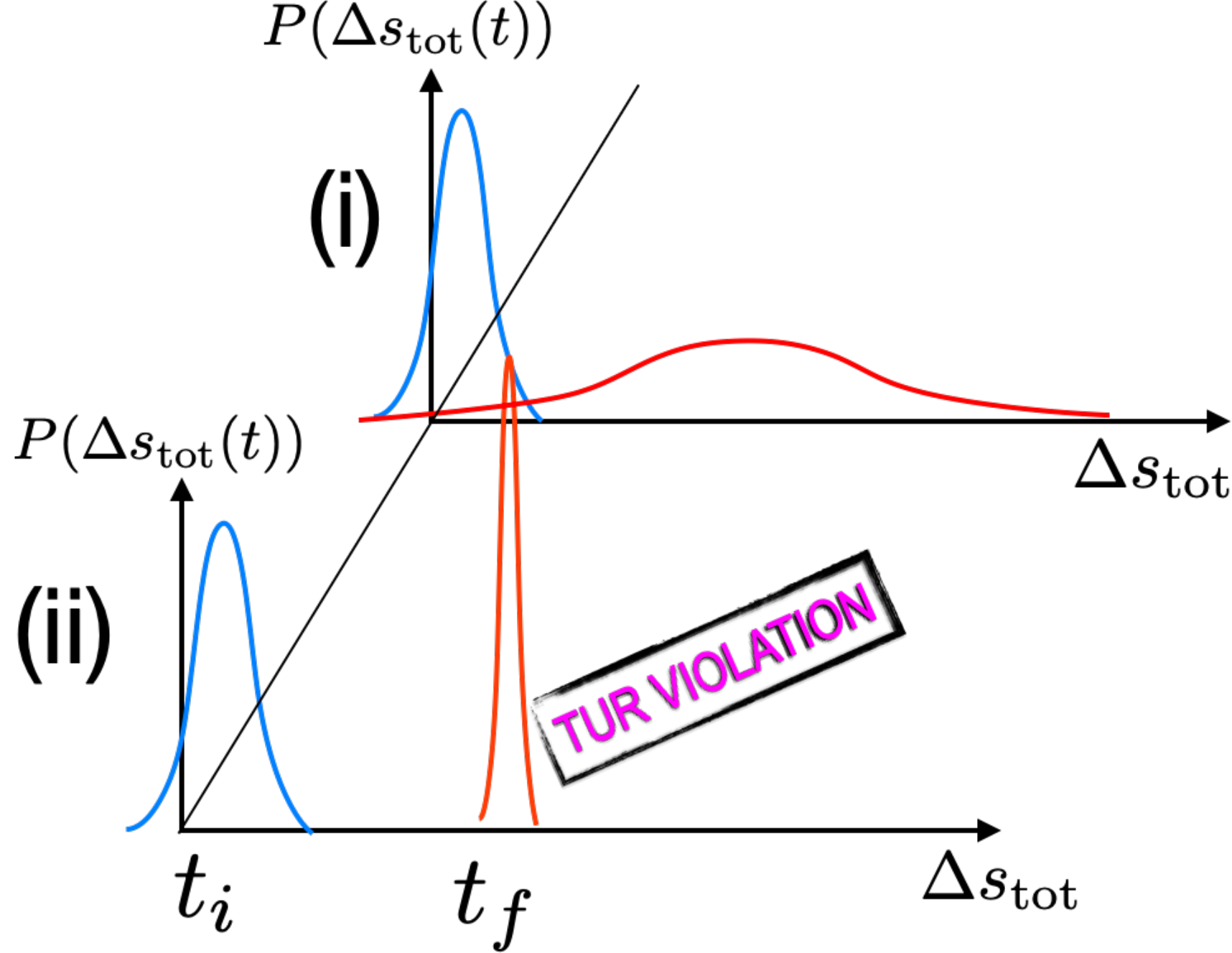}
	\caption{Properties of the distribution of total entropy production demanded by TUR. 
	Two scenarios of time evolution of $P(\Delta s_{\rm tot}(t))$ from $t=t_i$ (blue) to $t=t_f$ (red). 
	(i) As time increases, the distribution of entropy production is broadened. In this case, the TUR is obeyed with ${\rm Var}(\Delta s_{\rm tot})\geq 2\langle \Delta s_{\rm tot}\rangle$ at both $t=t_i$ and $t=t_f$. 
	(ii) As time increases, the distribution of entropy production is narrowed down, such that ${\rm Var}(\Delta s_{\rm tot})< 2\langle \Delta s_{\rm tot}\rangle$ at $t=t_f$, which contradicts to Eq.\ref{eq:TUR_sigma} and violates the TUR. 
	}  
\label{fig:TUR_distribution}
\end{figure}

If one chooses the trajectory-based entropy production (entropic current) as the output observable of interest such that $X(t)=\Delta s_{\rm tot}(t)$ \cite{Hyeon2017PRE,pietzonka2017PRE,manikandan2020PREL,dechant2020continuous},  
then the TUR in Eq.\ref{eq:TUR_original} can be written as 
\begin{align}
\mathcal{Q} =\frac{{\rm Var}(\Delta s_{\rm tot}(t))}{\langle \Delta s_{\rm tot}(t)\rangle}   \geq 2. 
\label{eq:TUR_sigma}
\end{align} 
Since the entropy production, an extensive variable, increases with time ($\Delta S_{\rm tot}=\langle \Delta s_{\rm tot}(t)\rangle \propto t$), the inequality between the variance and mean of entropy production, ${\rm Var}(\Delta s_{\rm tot}(t))\geq 2\langle \Delta s_{\rm tot}(t)\rangle$,  demands that the variance of entropy production also increases with $t$ but that it cannot be smaller than twice the mean entropy production (see Fig.\ref{fig:TUR_distribution}).  

Meanwhile, nonequilibrium dynamical processes discussed in this article are expected to establish cyclic steady states as long as the time scale of observation is longer than a single cycle time. 
The total entropy $\Delta s_{\rm tot}(t)$ produced from such processes obeys the \emph{detailed fluctuation theorem} (DFT) \cite{Seifert2012RPP,seifert2005PRL}, 
\begin{align}
\frac{P(\Delta s_{\rm tot})}{P(-\Delta s_{\rm tot})}=e^{\Delta s_{\rm tot}}  
\end{align} 
where we drop the dependence of entropy production on $t$ for simplicity of the notation.   
The DFT follows straightforwardly from Eq.\ref{eqn:FB_ratio} and the odd parity of entropy production upon time-reversal ($\Delta s_{\rm tot}(\Gamma)=-\Delta s_{\rm tot}(\tilde{\Gamma})$) by evaluating 
$P(\Delta s_{\rm tot})=\int \mathcal{D}\Gamma\delta (\Delta s_{\rm tot}-\Delta s_{\rm tot}(\Gamma))P(\Gamma)$,  
and it relates the probability density functions of entropy production from an irreversible process for time $t$ with exponentiated total entropy production. Rearranging the terms and integrating with respect to $\Delta s_{\rm tot}$ over $-\infty<\Delta s_{\rm tot}<\infty$ 
give rise to the \emph{integral fluctuation theorem} (IFT), $\langle e^{-\Delta s_{\rm tot}}\rangle=1$, 
\begin{align}
\langle e^{-\Delta s_{\rm tot}}\rangle &=\int_{-\infty}^{\infty} d(\Delta s_{\rm tot}) e^{-\Delta s_{\rm tot}}P(\Delta s_{\rm tot})\nonumber\\
&=\int_{-\infty}^{\infty}d(\Delta s_{\rm tot})P(-\Delta s_{\rm tot})=1.  
\label{eq:IFT}  
\end{align}
Here, $\langle\ldots\rangle$ denotes the average taken over the distribution of total entropy production from the process, $P(\Delta s_{\rm tot})$, and the relation should hold for any dynamical process. 
However, a special relation follows if the distribution of the total entropy production ($\Delta s_{\rm tot}$) is a Gaussian, i.e., 
$P(\Delta s_{\rm tot})\sim \exp{\left[-\frac{(\Delta s_{\rm tot}-\langle \Delta s_{\rm tot}\rangle)^2}{2{\rm Var}(\Delta s_{\rm tot})}\right]}$  
\begin{align}
1&=\langle e^{-\Delta s_{\rm tot}}\rangle=\int_{-\infty}^{\infty} d(\Delta s_{\rm tot})e^{-\Delta s_{\rm tot}}P(\Delta s_{\rm tot})\nonumber\\
&=e^{-\langle \Delta s_{\rm tot}\rangle
+{\rm Var}(\Delta s_{\rm tot})/2}, 
\end{align}
from which the equality condition of Eq.\ref{eq:TUR_sigma} ${\rm Var}(\Delta s_{\rm tot})/\langle \Delta s_{\rm tot}\rangle=2$ is acquired. 

The above argument based on IFT and Gaussian distribution suggests that 
the equality condition of TUR ($\mathcal{Q}=2$) is attained when the distribution of entropy production follows precisely a Gaussian. 
Since the tail part of the distribution contributes to the evaluation of the exponentiated average, the distribution of the total entropy production $\Delta s_{\rm tot}$, the sum of $\Delta s_i$'s ($i=1,2,\ldots t$) which are the \emph{independently and identically distributed} (i.i.d.) random variables ($\Delta s_{\rm tot}=\sum_{i=1}^t\Delta s_i$), should be assessed using the large deviation theory, an extension of central limit theorem \cite{touchette2009PR,Gingrich2016PRL}. 
In order for $P(\Delta s_{\rm tot})$ to be a perfect Gaussian including the tail parts, the entropy production from each cycle $P(\Delta s_i)$ should also be a Gaussian, which is highly restrictive.  
Thus, the inequality ($\mathcal{Q}>2$), leading to ${\rm Var}(\Delta s_{\rm tot})>2\langle \Delta s_{\rm tot}\rangle$, arises when $P(\Delta s_{\rm tot})$ deviates from the Gaussianity.  
This condition of Gaussianity of the distribution of total entropy production is more general than the condition of detailed balance (DB) or equilibrium, 
which gives rise to 
the equality condition of TUR for Markov process (see Eqs.\ref{eqn:Q_N1} and \ref{eqn:Q_N2}).
However, even at out-of-equilibrium, $\mathcal{Q}$ could be minimized (or sub-optimized) to a small value. 
An illuminating far-from-equilibrium case that gives rise to the equality condition ($\mathcal{Q}=2$) is found in a strongly driven colloidal particle in periodic potentials with $\Delta s_{\rm tot}(t)$ being chosen as the output observable \cite{Hyeon2017PRE,speck2005JPA,Pigolotti2017PRL,speck2007EPL}. 

Second, we underscore the significance of TUR in light of the second law of thermodynamics. 
Using the definitions of entropy production rate 
$\sigma_{\rm tot}(t)=\Delta S_{\rm tot}(t)/t$, mean and variance of finite time  current, $\langle J\rangle =\langle X(t)\rangle/t$ and $\langle \delta J^2\rangle =\langle \delta X(t)^2\rangle/t$, 
one can rewrite the expression of TUR in Eq.\ref{eq:TUR_original} as follows. 
\begin{align}
\mathcal{Q}=\sigma_{\rm tot}(t)\frac{\langle \delta J^2\rangle}{\langle J\rangle^2}\geq 2. 
\end{align}
Rearranged in the following way, 
\begin{align}
\sigma_{\rm tot}(t)\geq B\langle J\rangle^2
\label{eq:Sasa}
\end{align}
with $B=\left(2/\langle\delta J^2\rangle\right)$, 
the TUR conveys a remarkable message that the rate of entropy production from the dynamical process is bounded below by the square of mean current multiplied by a prefactor $B$.
This makes the physical bound of thermodynamic process more explicit and tighter than the second law of thermodynamics ($\Delta S_{\rm tot}\geq 0$ or $\sigma_{\rm tot}(t)\geq 0$) \cite{dechant2018JSM,dechant2018PRE,horowitz2019NaturePhys}.  
Recently, Li \emph{et al.}\cite{li2019NatComm} have shown that once inferring the macroscopic current and its fluctuations, the TUR can be used to estimate the lower bound of the entropy production (dissipation) rate even for high-dimensional dynamical systems.

\section*{The uncertainty product $\mathcal{Q}$ as a measure of optimality of enzymatic processes}
Many biological processes, which exhibit a net change in terms of chemical species, are driven by enzyme reactions that incur thermodynamic cost. 
To discuss the TUR in the context of the enzyme reactions, 
it is convenient to recast the expression (Eq.~\ref{eq:TUR_original}) into 
\begin{align}
\mathcal{Q} =  \frac{\langle \Delta s_{\rm tot}(t)\rangle}{\langle n(t) \rangle}\frac{{\rm Var}(n(t))}{\langle n(t) \rangle} = \beta\mathcal{A}\frac{\langle \delta J^2 \rangle}{\langle J\rangle} = \beta\mathcal{A}\times \lambda \geq 2,
\label{eq:enzyme_Q}
\end{align}
where we choose $n(t)$ the net number of catalyses that take place for time $t$ as the observable of interest, defining the mean and fluctuation of finite time current as $\langle J\rangle = \langle n(t)\rangle/t$, $\langle \delta J^2\rangle = {\rm Var}(n(t))/t$.  
The Fano factor associated with the number of catalyses or with the reaction current is defined as $\lambda ={\rm Var}(n(t))/\langle n(t)\rangle=\langle \delta J^2 \rangle/{\langle J\rangle}$, which corresponds to the randomness parameter \cite{schnitzer1995statistical,Visscher99Nature}, and 
$\beta\mathcal{A}= \langle \Delta s_{\rm tot}(t)\rangle/\langle n(t)\rangle$ with the inverse temperature $\beta= (k_BT)^{-1}$
is the entropy production per cycle (or affinity) of the enzymatic reaction. 
Since $\langle J\rangle$ and $\langle \delta J^2\rangle$ are associated with the speed ($\langle J\rangle\sim Vt$) and diffusivity ($\langle \delta J^2\rangle\sim 2Dt$) of the enzymatic process, 
a dynamic process that operates at high speed, low fluctuation (high regularity), and low thermodynamic cost is characterized with a small $\mathcal{Q}$ and can be deemed more optimally designed when $\mathcal{Q}$ is smaller and close to 2 \cite{dechant2018JSM,Hwang2018JPCL}. 

\begin{figure}[t]
	\includegraphics[width=1\linewidth]{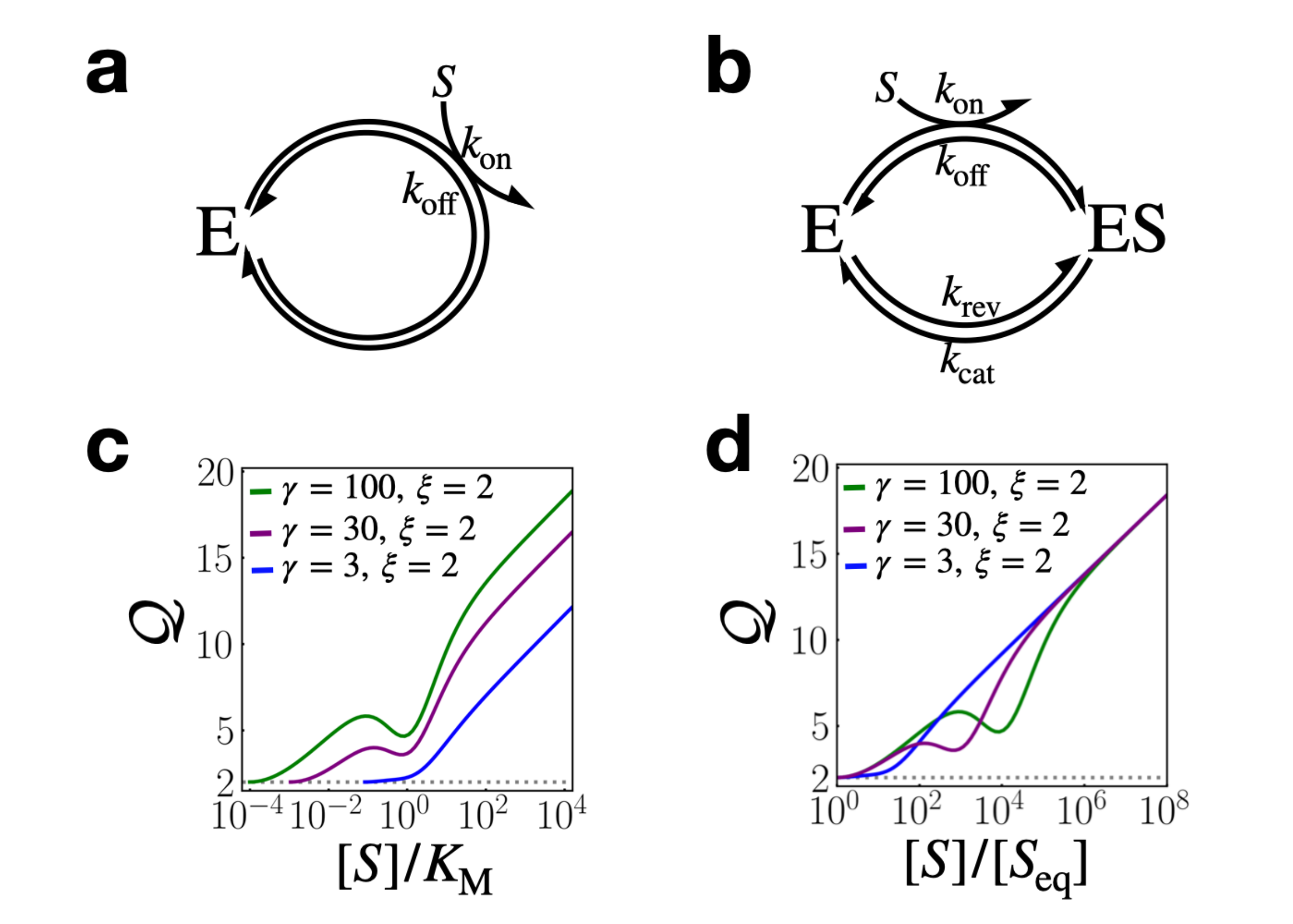}
	\caption{ $\mathcal{Q}$ of simple enzyme reactions as a function of the substrate concentration $[S]$. 
	Schematics of (a) the 1-state and (b) 2-state kinetic networks. 
	(c)(d) $\mathcal{Q}([S])$ of the 2-state kinetic network, for three pairs of $\gamma$ and $\xi$ values. 
	In (c), $[S]$, normalized by $K_\text{M}$, shows that 
	$\mathcal{Q}([S])$ is locally minimized at $[S]\approx K_{\rm M}$.
	In (d), $[S]$ is normalized by $[S]_{\rm eq}[=(k_{\rm off}k_{\rm rev})/(k_{\rm on}k_{\rm cat})]$, the substrate concentration at the detailed balance condition, showing that $\lim_{[S]\rightarrow [S]_\text{eq}}\mathcal{Q}([S]) = 2$.
	The rate constants used for the plots are as follows: $k_{\rm on}=10^8$ M$^{-1}$sec$^{-1}$, $k_{\rm off}=k_{\rm rev}=10$ sec$^{-1}$, and $k_{\rm cat}=30, 300, 1000$ sec$^{-1}$.
	}
	\label{fig:Fig2}
\end{figure}

Before exploring physically realistic processes, such as molecular motors and biomass producing machineries, 
we motivate by computing $\mathcal{Q}$ of two simple versions of enzymatic cycle below. 

(i) For the catalysis of the substrate $S$ in the ($N=1$)-state kinetic reaction network (Fig.\ref{fig:Fig2}a), the binding and unbinding rates of the substrate are denoted by 
$k_{\rm on}$ and $k_{\rm off}$, respectively. 
The substrate concentration, $[S]$, 
can be varied independently as a control parameter. 

The master equation for the probability $P_n(t)$ of observing $n$ catalytic events at time $t$ is 
\begin{align}
\dot{P}_n(t) = k_{\rm on}[S] P_{n-1}(t) + k_{\rm off} P_{n+1}(t) - (k_{\rm on}[S] + k_{\rm off})P_n(t),
\label{eqn:master}
\end{align}
with the initial condition $P_n(0) = \delta_{n,0}$. 
The method of generating function \cite{vanKampen} $F(z,t) =\sum_{n=-\infty}^{\infty} z^nP_n(t)$ casts Eq.\ref{eqn:master} into 
$F(z,t)=\exp{\left[zk_{\rm on}[S]+k_{\rm off}/z-(k_{\rm on}[S] + k_{\rm off})\right]}$, which allows one to calculate 
the mean and fluctuations in the number of catalytic events using  
$\partial_z F(z,t)\big|_{z=1} = \langle n(t)\rangle=(k_{\rm on}[S]-k_{\rm off})t$
and $\partial_z^2F(z,t)\big|_{z=1}= \langle n^2(t)\rangle -\langle n(t) \rangle=2k_{\rm off}t+(k_{\rm on}[S]-k_{\rm off})^2t^2$.
Thus, the quantities required to evaluate $\mathcal{Q}$ in Eq.\ref{eq:enzyme_Q} are obtained as follows. 
\begin{align}
 \langle J \rangle &= k_{\rm on}[S]-k_{\rm off}, 
 \end{align}
 \begin{align}
 \langle \delta J^2\rangle&=k_{\rm on}[S]+k_{\rm off}, 
 \end{align}
 \begin{align}
\lambda &= \frac{\langle \delta J^2\rangle}{\langle J \rangle} = \frac{e^{\beta \mathcal{A}}+1}{e^{\beta \mathcal{A}}-1}, 
\end{align}
 \begin{align}
\beta\mathcal{A}=&\ln\frac{k_{\rm on}[S]}{k_{\rm off}}, 
\end{align}
which leads to 
\begin{align}
\mathcal{Q} =& \beta\mathcal{A}\frac{e^{\beta \mathcal{A}}+1}{e^{\beta \mathcal{A}}-1}\geq 2. 
\label{eqn:Q_N1}
\end{align}
At the DB condition ($[S]=[S]_{\rm eq}\equiv k_{\rm off}/k_{\rm on}$), 
no net catalysis ($\langle J\rangle=0$) and no dissipation ($\mathcal{A}=0$) occurs, 
and the uncertainty product reaches its bound, $\mathcal{Q} = 2$. 
As $[S]$ increases, breaking the DB condition,  $\langle J\rangle$, $\mathcal{A}$, $\lambda$, and $\mathcal{Q}$ all increase monotonically.

(ii) For substrate catalysis through the ($N=2$)-state kinetic network, with the rate constants as denoted in Fig.~\ref{fig:Fig2}b, 
we obtain the following expressions 
\cite{Fisher01PNAS,KolomeiskyBook,Hwang2017JPCL} 
for the quantities required to evaluate $\mathcal{Q}$:
\begin{align}
\langle J \rangle&= \frac{k_{\rm on}k_{\rm cat}[S]-k_{\rm rev}k_{\rm off}}{k_{\rm on}[S]+k_{\rm cat}+k_{\rm rev}+k_{\rm off}} 
\label{eq:J}
\end{align}
\begin{align}
\langle \delta J^2 \rangle&=  \frac{  
k_{\text{on}}k_{\text{cat}}[S] + k_{\text{rev}}k_{\text{off}}
-2 \langle J \rangle^2
  }{k_{\rm on}[S]+k_{\rm cat}+k_{\rm rev}+k_{\rm off}} 
  \label{eq:J2}
  \end{align}
  \begin{align}
\lambda &=\frac{k_{\rm on}k_{\rm cat}[S]+k_{\rm rev}k_{\rm off}}{k_{\rm on}k_{\rm cat}[S]-k_{\rm rev}k_{\rm off}} -2\frac{k_\text{on}k_{\text{cat}}[S]-k_\text{rev}k_\text{off}}{k_\text{on}^2\left([S]+K_{\rm M}+\frac{k_\text{rev}}{k_\text{on}}\right)^2}
\label{eq:lambda2} 
\end{align}
\begin{align}
\beta\mathcal{A}&=\ln{\frac{k_{\rm on}k_{\rm cat}[S]}{k_{\rm rev}k_{\rm off}}} 
\label{eq:A}
\end{align}
with the Michaelis constant, $K_{\rm M}=(k_\text{cat}+k_\text{off})/k_\text{on}$. 
When $\mathcal{Q}$ is written as a function of the thermodynamic drive $\beta\mathcal{A}$, 
\begin{align} 
\mathcal{Q}  &= \beta\mathcal{A}\left[ \frac{e^{\beta \mathcal{A}}+1}{e^{\beta \mathcal{A}}-1} 
- 2\frac{ \gamma^2 \left(  e^{\beta \mathcal{A}}-1 \right) }{ \left( \gamma^2+\gamma\xi +e^{\beta \mathcal{A}} \right)^2 }\right], 
\label{eqn:Q_N2}
\end{align}
with dimensionless parameters $\gamma = \frac{k_{\text{cat}}}{\sqrt{{k_{\text{off}}}{k_{\rm rev}}}}$ and $\xi = \frac{k_{\text{off}}+k_{\rm rev}}{\sqrt{k_{\text{off}}k_{\rm rev}}}(\geq 2)$, 
one can show that for any $\xi(\geq 2)$ there exists a threshold value of $\gamma$, above which $\mathcal{Q}$ becomes non-monotonic with $[S]$ (see Fig.~\ref{fig:Fig2}c, and supplementary Fig.~2 in \cite{Song2020JPCL}). 
One can also show that for $\beta\mathcal{A}\ll 1$, 
$\mathcal{Q}=2+\left[\frac{1}{6}-\frac{2\gamma^2}{(1+\gamma\xi+\gamma^2)^2}\right](\beta\mathcal{A})^2+\mathcal{O}\left[(\beta\mathcal{A})^3\right]\geq 2$, which confirms that $\mathcal{Q}=2$ is attained at DB condition (Fig.~\ref{fig:Fig2}d).  
Additionally, it is straightforward to see from Eq.~\ref{eq:lambda2} that,
at the limit of a strongly driven process with $\beta\mathcal{A}\gg 0$ and $k_\text{on}\gg k_\text{rev}$, 
$\lambda$ is  simplified to 
\begin{align}
\lambda([S]) \simeq 1-\frac{2k_{\text{cat}}[S]}{k_{\text{on}}([S]+K_\text{M})^2},
\end{align} 
which is minimized at $[S]={ K_\text{M}}$ (Fig.~\ref{fig:Fig2}c). 
Thus, when $k_{\rm on}k_{\rm cat}[S]\gg k_{\rm rev}k_{\rm off}$ ($\beta\mathcal{A}\gg 0$) and $k_{\text{on}} \gg k_{\rm rev}$,
$\mathcal{Q}(\text{[S]})$ has a local minimum and is sub-optimized at around $[\text{S}] \approx { K_\text{M}}$ (Fig.~\ref{fig:Fig2}c).

\begin{figure*}[t]
	\includegraphics[width=1\linewidth]{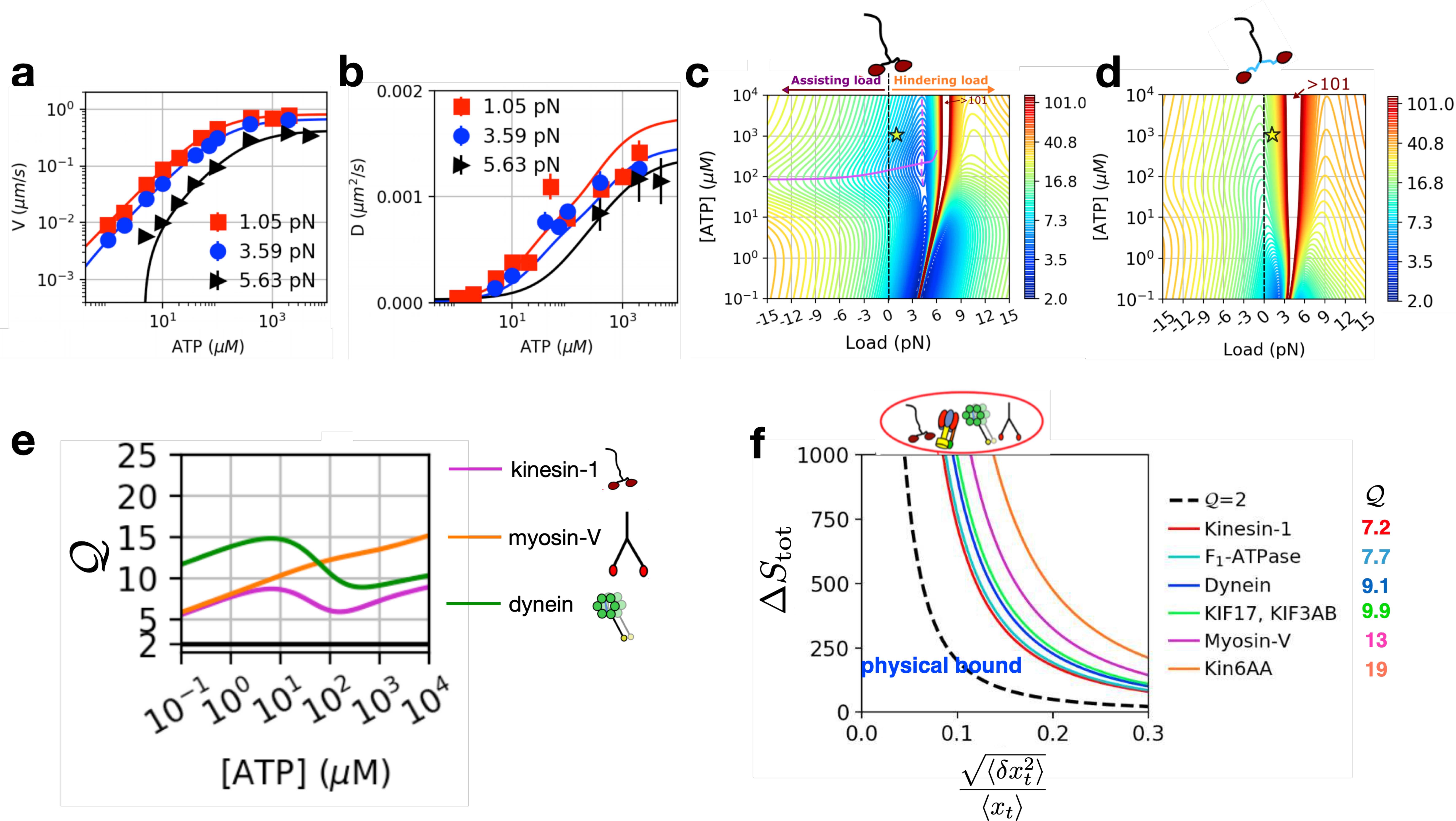}
	\caption{
	 \textbf{(a)(b)} Analysis of experimental data of (a) $V$ and (b) $D$ of kinesin-1.
	 The experimental data are digitized from Ref.~\cite{Visscher1999}, and fit using a 6-state double-cycle kinetic model in Ref.~\cite{Hwang2018JPCL}. The solid lines are fits to the data, with three values of external load. 
	 \textbf{(c)(d)} Diagrams of $\mathcal{Q}$ as a function of [ATP] and external load ($f$). 
	 Assisting and resisting loads correspond to $f<0$ and $f>0$, respectively;
	 {(c)} kinesin-1,
	 {(d)} kinesin-1 mutant. 
	 The color code and the corresponding colorbar in (c)(d) quantify the value of the uncertainty product $\mathcal{Q}$. 
	 \textbf{(e)} The values of $\mathcal{Q}$ as a function of ATP concentration at $f=1$ pN. 
	 \textbf{(f)} The tradeoff relations between $\Delta S_{\rm tot}$ and the relative uncertainty of displacement $x_t(\equiv x(t))$ are plotted for various molecular motors with their uncertainty products calculated at $[{\rm ATP}]=1$ mM and $f=1$ pN. 
	 The figure was adapted from Ref.\cite{Hwang2018JPCL}.}
	\label{fig:Q_diagram}
\end{figure*}

The non-monotonic variations of $\mathcal{Q}([S])$ seen in some of the biological motors and copy enzymes discussed below simply means that the dynamical processes of these molecules display a Michaelis-Menten type hyperbolic dependence on $[S]$. 
The condition of $[S]\approx K_\text{M}$ in the Michaelis-Menten type reaction mechanism, $v([S])=k_{\rm cat}[S]/(K_\text{M}+[S])$, corresponds to the point where the response of enzymatic activity to the logarithmic change in substrate concentration is maximized, i.e., $dv/d\log{[S]}=k_{\rm cat}(K_\text{M}/[S])/(1+K_\text{M}/[S])^2\leq k_{\rm cat}/4$. 
Importantly, a recent systems level analysis of metabolic pathways in eukaryotic cells has shown that the physiological substrate concentrations of many enzyme reactions are tuned near their respective $K_\text{M}$ values \cite{Park2016}, which naturally explains the suboptimality of  $\mathcal{Q}([S])$ at $[S]\approx K_\text{M}$ for some processes discussed in the following sections. 
\\

\section*{Transport processes by molecular motors}
Biological motors are a class of enzymes that have an ability to transduce chemical free energy to mechanical motion via the catalysis of molecular fuels (ATP, GTP, etc) present in the cellular milieu \cite{KolomeiskyBook,howard2001mechanics}. 
Among them, kinesin-1 is arguably the most well-studied motor protein that transports cargos or organelles from the minus to plus ends of microtubules with the velocity of $V\approx 1$ $\mu$m/sec, taking 8 nm step for every ATP hydrolysis. 
Dynein moves with a similar speed, but in the opposite direction along microtubules, displaying more fluctuations in the time traces. 
Myosin families are the motor proteins specialized to move along actin filaments and generate mechanical forces. 
Whereas these motor proteins are specialized for linear movement and force generation, 
there are also rotary motors (e.g., F$_1$F$_0$ATPase) that utilize $H^+$ gradient across membrane to generate rotational motion (or torque), which is, for example, used to empower the beating dynamics of bacterial flagella. 

The chemical free energy-driven dynamics of molecular motors are a perfect example whose optimality can be assessed in light of TUR and the uncertainty product $\mathcal{Q}$. 
The TUR in the original form (Eq.\ref{eq:TUR_original}) with $X(t)=\Delta x(t)$ can be cast into the following form with $V=\langle \Delta x(t)\rangle /t$ and $D=[\langle \Delta x(t)^2\rangle -\langle \Delta x(t)\rangle^2]/2t$, 
\begin{align}
\mathcal{Q}=\sigma_{\rm tot}\frac{2D}{V^2}\geq 2.  
\label{eq:motor_Q}
\end{align}
According to this relation, a molecular motor that transports cargos with high velocity ($V$), low fluctuations ($D$), and low thermodynamic cost ($\sigma_{\rm tot}$) would be characterized with a small value of $\mathcal{Q}$, 
and one could argue that a motor with smaller $\mathcal{Q}$ value is 
better designed as a molecular transporter. 
For a \emph{transport efficiency} defined as $\eta_{\rm tr}\equiv 2/\mathcal{Q}$, which is bounded in $0\leq \eta_{\rm tr}\leq 1$,  
a motor with smaller $\mathcal{Q}$ can be said to have a higher \emph{transport efficiency} \cite{dechant2018JSM,Hwang2018JPCL}. 

Of note, the TUR can also be used to determine the upper bounds on the \emph{thermodynamic efficiency} by means of
the dissipation bound, $\sigma_{\rm tot}\geq V^2/D$ (Eq.\ref{eq:motor_Q}).  
Pietzonka \emph{et al.}\cite{pietzonka2016JSM,uhl2018PRE} 
defined the \emph{thermodynamic efficiency} ($\eta$) of a molecular motor in the presence of external load ($f$)
as the ratio between the amount of work production ($\dot{W}=fV$) and an input chemical potential ($-\dot{\Delta\mu}=\dot{W}+T\sigma_{\rm tot}$), so as to obtain its upper bound of $\eta$ as follows. 
\begin{align}
\eta&=\frac{\dot{W}}{|\dot{\Delta\mu}|}
=\frac{\dot{W}}{\dot{W}+T\sigma_{\rm tot}}=\frac{1}{1+\frac{TV}{2fD}\mathcal{Q}}\nonumber\\
&\leq \frac{1}{1+\frac{TV}{fD}} =\eta_{\rm max}. 
\end{align}
The maximum thermodynamic efficiency, $\eta_{\rm max}$, is obtained in the condition equivalent to $\mathcal{Q}=2$.

Given the single molecule time traces, $\{x(t)\}$, the velocity ($V$) and diffusivity ($D$) are straightforwardly determined. 
Meanwhile, the rate of total entropy production $\sigma_{\rm tot}$ either (i) can be estimated, for example, from the fact that each step of kinesin-1, which occurs every $\sim 10$ msec, is generated from the hydrolysis free energy of a single ATP molecule  $\sim$ $20$ $k_BT$ ($T\sigma_{\rm tot}\sim 20$ $k_BT/10 \text{  msec}$), or (ii) can be calculated more systematically by knowing all the chemical rate constants on the kinetic network \cite{Schnakenberg1976RMP}.  
$\sigma_{\rm tot}$ for ($N=2$)-unicyclic network in the absence of external load ($f=0$), for example, can be calculated using Eqs.\ref{eq:J}, \ref{eq:A}
\begin{align}
\sigma_{\rm tot}&=\langle J\rangle\times \beta\mathcal{A}\nonumber\\
&=\frac{k_{\rm on}k_{\rm cat}[S]-k_{\rm rev}k_{\rm off}}{k_{\rm on}[S]+k_{\rm cat}+k_{\rm rev}+k_{\rm off}}\log{\left(\frac{k_{\rm on}k_{\rm cat}[S]}{k_{\rm rev}k_{\rm off}}\right)}. 
\end{align}
For the case (ii), $\sigma_{\rm tot}$ of a general kinetic network with known transition rate constants, can be obtained by a method developed by Koza \cite{Koza1999}. 

\begin{figure*}[t]
	\includegraphics[width=1.0\linewidth]{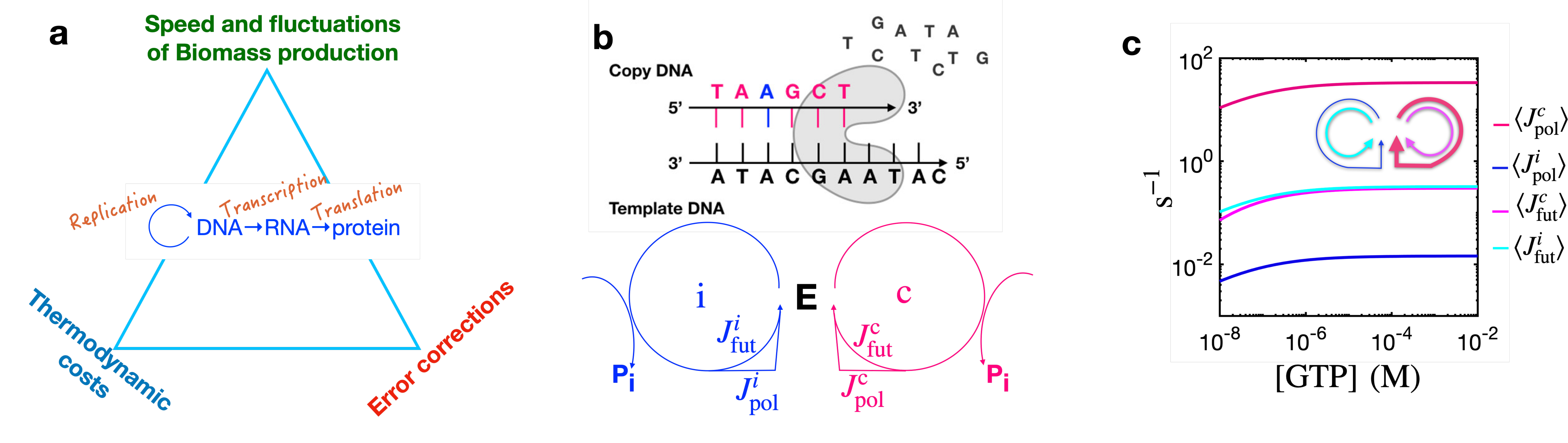}
	\caption{Biological copy processes and tradeoff relation. 
	\textbf{(a)}  Tradeoff relations in biological copy processes. 
	Excessive amount of error corrections are expected to lead to higher thermodynamic cost and slow biomass production. 
	\textbf{(b)} The illustration of DNA replication process (top), and a kinetic network to represent general copy processes with the possibility of accommodating either correct or incorrect substrate to copy strand.  
	The $c$-cycle (red) and $i$-cycle (blue) are for the correct and incorrect substrate incorporations, respectively. 
	\textbf{(c)} Reaction currents along the subcycles of the network as a function of substrate concentration. 
	Explicit calculations of currents were conducted using an example of translation of a codon CUG by ribosome \cite{Song2020JPCL}. 
	The figures b and c were adapted from Ref.\cite{Song2020JPCL}.
	}
	\label{fig:copy_process}
\end{figure*}

As clearly gleaned from the data extracted from single molecule time traces of a molecular motor (e.g., kinesin-1), the velocity and diffusivity of time traces 
vary with ATP concentration ([ATP]) and external force ($f$) (Fig.\ref{fig:Q_diagram}a,b) \cite{Schnitzer97Nature,Visscher99Nature,Schnitzer00NCB}. 
Thus, the value of $\mathcal{Q}$ not only depends on a type of  molecular process of interest, but also on its working condition. 
Fig.~\ref{fig:Q_diagram}c and d shows the diagrams of $\mathcal{Q}$ as a function of $f$ and [ATP] for kinesin-1  and kinesin-1 mutant, respectively. 
Some features of $\mathcal{Q}(f,[{\rm ATP}])$ of biological motors are noteworthy: 
(i) As clearly seen for the case of kinesin-1 (Fig.\ref{fig:Q_diagram}c), besides the trivial minima at low [ATP] corresponding to DB condition, $\mathcal{Q}$ is sub-optimized at $f\approx 3$ pN and $[{\rm ATP}]\approx 200$ $\mu M$. 
$\mathcal{Q}(f=1 {\rm pN}, [{\rm ATP}])$ plotted in Fig.\ref{fig:Q_diagram}e displays non-monotonic variation with [ATP]. A similar behavior is observed in $\mathcal{Q}$ for dynein (Fig.4 of ref.~\cite{Hwang2018JPCL}). 
As already discussed in the foregoing section, the non-monotonic variation of $\mathcal{Q}([{\rm ATP}])$ is the outcome of Michaelis-Menten type response of motor or enzymatic activity on substrate concentration. 
(ii) The cellular condition ($f\approx 1$ pN, [ATP] $\approx 1$ mM) marked with a yellow star (Fig.\ref{fig:Q_diagram}c), is in proximity to the suboptimal condition. 
(iii) The high $\mathcal{Q}$ value region in the middle of the diagram ($\mathcal{Q}>100$) 
arises when the hindering load stalls the movement of motor. 
It is important to note that at the stall condition ATP is still consumed ($\sigma_{\rm tot}>0$) \cite{Cross05Nature,Hyeon09PCCP,Hwang2018JPCL,sumi2019NanoLett}. 
While the motor is motionless at mechanical equilibrium ($V\approx 0$), it is not in chemical equilibrium ($\sigma_{\rm tot}>0$). As a result, $\mathcal{Q}$ diverges (Eq.\ref{eq:motor_Q}) 
(see the high $\mathcal{Q}$ region in Figs.\ref{fig:Q_diagram}c and \ref{fig:Q_diagram}d). 
(iv) $\mathcal{Q}(f,[{\rm ATP}])$ of kinesin-1 mutant (Fig.\ref{fig:Q_diagram}d), which has six additional amino acids inserted into neck-linker, is drastically altered from that of the wild-type, such that the value of $\mathcal{Q}$ is increased and the stall condition is formed at smaller $f$ values. 
(v) Lastly, Fig.\ref{fig:Q_diagram}f compares the values of $\mathcal{Q}$ for various biological motors at the cellular condition ($f\approx 1$ pN, [ATP] $\approx 1$ mM). 
$\mathcal{Q}=7-15$ for wild type motors: kinesin-1, F$_1$-ATPase, dynein, KIF17, KIF3AB, and myosin-V. 
In particular, $\mathcal{Q}=7$ for kinesin-1, whereas its mutant displays much deteriorated performance with $\mathcal{Q}=19$. 
The uncertainty product of biological motor specialized for cargo/organelle transport is relatively smaller than other machineries that will be discussed below.    
\\

\section*{Biological copy processes}
As exemplified in DNA replication, transcription, and translation processes which comprise the central dogma of molecular biology, some of the key information transfer processes in biology are nearly free from copy error even in the noisy cellular environment.
For the case of DNA replication, the error probability ($\eta_{eq}$) estimated solely from the stability difference between the correct and incorrect base pairs is at best $\eta_{eq}\approx 10^{-3}$ ~\cite{Goodman1997}; however, the actual error probability of replicating incorrect bases to copy strand is as small as $\eta=10^{-10}$, which makes the replication of giant human DNA consisting of $N=3\times 10^9$ bases effectively error-free \cite{Kunkel2000}. 
To achieve the substantial error reduction from $\eta_{eq}$ to $\eta$, a host of elaborate energy-expending molecular mechanisms are at work at every step of the biological copy processes \cite{Ibarra2009,Shaevitz2003,Blanchard2004,Cvetesic2012}. 

From the viewpoint of information processing in biology, 
the error reduction is certainly an important issue; however, it in itself cannot be the sole goal of the biological copy processes given that biomass production, e.g., from DNA to RNA, and from RNA to proteins, is the major outcome from the processes. 
Excessive operations of error correction machineries would not only incur thermodynamic cost, but also slow down the process of biomass production. 
There have recently been a number of studies 
devoted to understanding the strategy of biological copy processes to achieve the mutually contradicting goals of low copy error, high speed, and low thermodynamic cost \cite{Murugan2012a,Gaspard2016b,Gaspard2016a,Banerjee2017,Wong2018,Mallory2019} (Fig.~\ref{fig:copy_process}a).

Here, we will view each step of copying enzyme along a template polymer as that of a molecular motor stepping along a template filament processively. 
Each substrate incorporating event 
can be split into several Markov jump steps on a cyclic kinetic network,
at the end of which the copying enzyme transitions to the next position along the template strand.
The forward motion of the enzyme is driven by the chemical potential of the biosynthetic substrates (dNTPs for the DNA polymerase, NTPs for the RNA polymerase, and charged tRNAs for the ribosome). 
Under an assumption that cellular homeostasis maintains the substrate concentrations constant but away from the DB condition, 
the dynamical process associated with synthesizing a copy strand can be modeled as a process operating at NESS with the substrate concentration, $[S]$, as the key parameter to be controlled. 

Instead of the uni-cyclic kinetic network introduced earlier for enzymatic processes on a single type of substrate catalysis, 
additional cycles due to the chance of copying incorrect types of substrate to the strand are necessary for the copy processes described above. 
The schematic of the reaction network in Fig.\ref{fig:copy_process}b represents the one for copy processes, which takes into account the chance of incorporating incorrect substrates. 
The $c$ denotes the reaction cycle (red) associated with \emph{correct} substrate incorporation, whereas $i$ is for the cycle (blue) associated with \emph{incorrect} substrate incorporation. 
Provided that correct substrates are always accommodated to the copy enzyme without any futile attempt,  
the reaction current, more specifically the current associated with polymerization, $J^c_{\text{pol}}$, flows only through the $c$-cycle.  
In this case, the probability of copy error would be 0 (Eq.\ref{eq:error_prob}, see below).  
However, the stochastic nature of biochemical reactions makes the incorporation of incorrect substrate to the copy polymer still inevitable. 
An incorrect substrate incorporated to the enzyme-cofactor complex engenders the reaction current through $i$-cycle. 
A proofreading mechanism expending the free energy (e.g., GTP hydrolysis for the case of mRNA translation by ribosome) 
in action filters out the incorrect substrate from the system, generating a futile current ($J^i_\text{fut}$); otherwise the substrate is accommodated into the copy polymer, generating a current associated with polymerization ($J^i_\text{pol}$).   
The futile current along the $c$-cycle is conceivable as well, just like the case in which a molecular motor, say kinesin-1, occasionally fails to step in spite of ATP hydrolysis. 
Explicit calculations of four currents in Fig.~\ref{fig:copy_process}b conducted for the translation of CUG codon by ribosome indicate that the sizes of the four currents are maintained in the following order, effectively over the whole range of GTP concentration (1 nM $\leq$ [GTP] $\leq$100 mM)\cite{Song2020JPCL}, while the cellular concentration of GTP is [GTP] $\approx 5$ mM \cite{Bennett2009}:  
\begin{align}
\langle J_\text{pol}^c \rangle\gg\langle J_\text{fut}^i \rangle\gtrsim \langle J_\text{fut}^c \rangle\gg \langle J_\text{pol}^i \rangle.  
\end{align}
Both $\langle J_{\rm pol}^c\rangle$ and $\langle J_{\rm pol}^i\rangle$ are reflected to the 
copied sequence, such that the error probability of the copy process is associated with the two mean currents as 
\begin{align}
\eta =\frac{\langle J^{i}_{\rm pol} \rangle}{\langle J^{c}_{\rm pol} \rangle+\langle J^{i}_{\rm pol} \rangle}, 
\label{eq:error_prob}
\end{align}
which gives $\eta\approx 4\times 10^{-4}$ for the case of CUG codon \cite{Song2020JPCL}.  

\begin{figure*}[t]
	\includegraphics[width=.9\linewidth]{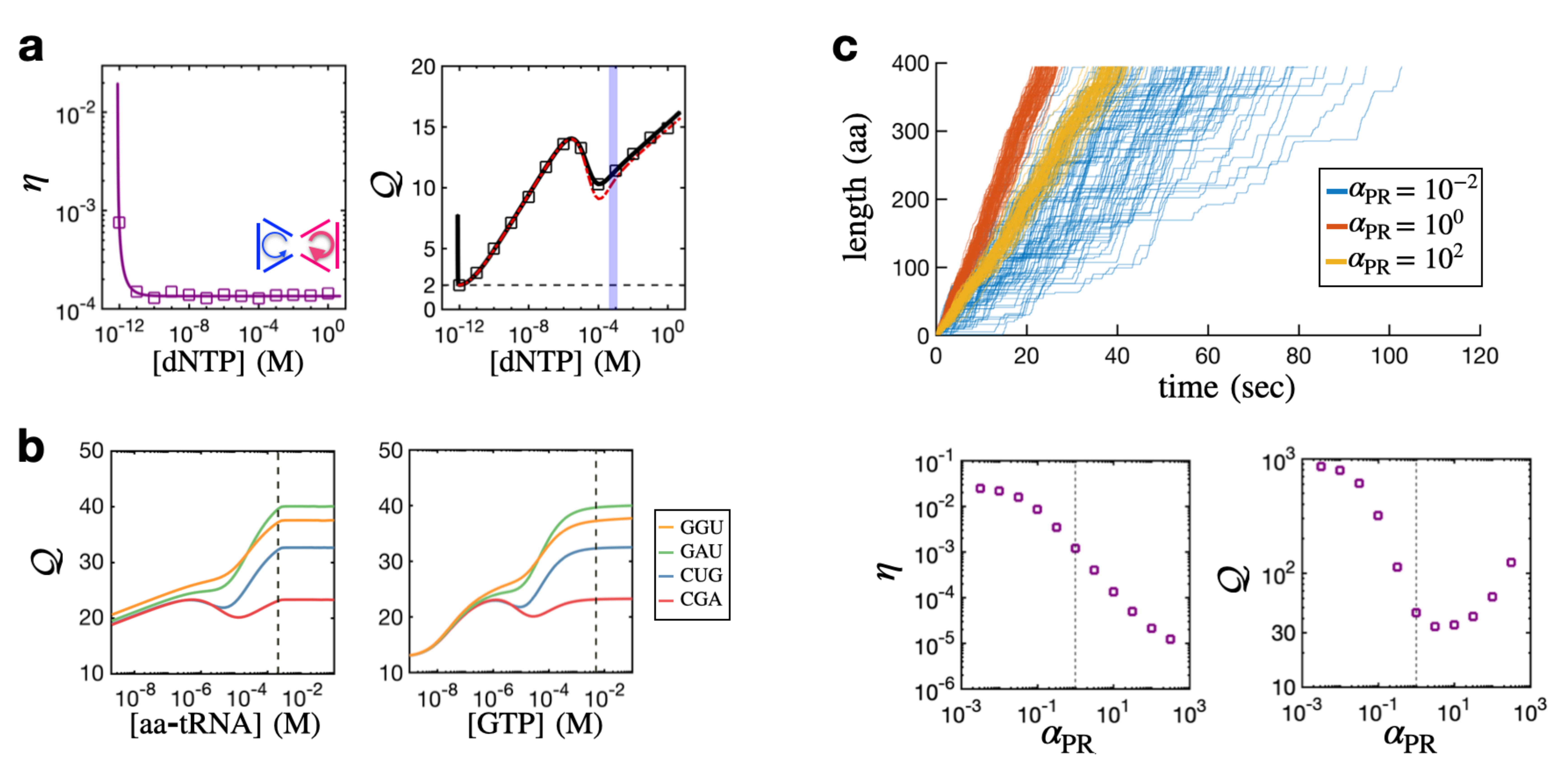}
	\caption{The uncertainty product $\mathcal{Q}$ for biological copy processes. {\bf (a)} DNA replication by T7 DNA polymerase. 
	{\bf (b)} Translation of codons by \emph{E. coli.} ribosome.  
	The dashed lines indicate the cellular concentrations of aa-tRNA and GTP. 
	{\bf (c)} Proofreading step-modulated fluctuations in the translation times of \emph{E. coli.} ribosome ($\mathcal{T}$) that reads the \emph{tufB} mRNA sequence encoding the 394 amino acids of EF-Tu. 
	In the two panels at the bottom are shown the error probability and uncertainty product as a function of the perturbation factor ($\alpha_{\rm PR}$) multiplied to the rate of proofreading step. 
	The figure was adapted from Ref.\cite{Song2020JPCL}.
	}
	\label{fig:Q_copy}
\end{figure*}

The free energy cost associated with the copy process represented by the schematic of kinetic network in Fig.\ref{fig:copy_process}b is given as follows \cite{Song2020JPCL}. 
\begin{align}
\beta\mathcal{A} =&
-\beta \left[ \Delta\mu_{\text{pol}} + \frac{ \langle J_{\rm fut} \rangle}{\langle J_{\text{pol}} \rangle}\Delta\mu_{\rm fut} \right]
-\eta \ln{\eta} - (1-\eta) \ln{(1-\eta)}. 
\label{eqn:affinity}
\end{align}
where $\langle J_{\text{pol}} \rangle \equiv \langle J^c_{\text{pol}} \rangle+\langle J^i_{\text{pol}} \rangle$ 
and $\langle J_{\text{fut}} \rangle \equiv \langle J^c_{\text{fut}} \rangle+\langle J^i_{\text{fut}} \rangle$ 
are the total reaction current along the polymerization and futile cycles.  
Note that since $\Delta\mu_\text{pol}$ and $\Delta\mu_\text{fut}$, the chemical potentials associated with each cycle, are the state function, they are identical regardless of the cycle.  
The value of $\beta\mathcal{A}$ is determined by the ratio of currents 
$\langle J_{\rm fut}\rangle/\langle J_{\rm pol}\rangle$ as given in the expression of Eq.\ref{eqn:affinity}. 
Lastly, the Shannon entropy-like term, $S(\eta)=-\eta \ln{\eta} - (1-\eta) \ln{(1-\eta)}$ 
arises from the entropic drive created by a potential disorder in copied sequence.  
Although the actual contribution of $S(\eta)$ is minor compared to the entire free energy ($\mathcal{A}$) in biologically relevant parameter regimes \cite{Song2020JPCL}, 
$S(\eta)$ could, in principle, be the source of thermodynamic drive for the polymerization reaction when $[\Delta\mu_{\rm pol} + (\langle J_{\rm fut} \rangle/\langle J_{\rm pol}\rangle)\Delta\mu_{\rm fut} ]\approx 0$ (see Ref. \cite{bennett1982IJTP}). 
Eq.~\ref{eqn:affinity} clarifies that extra free energy cost is incurred in the presence of a larger amount of futile current, when the proofreading  mechanisms are at work.

Finally, we aim to look at the problem of biological copy processes under the hood of TUR and study how the error probability is balanced with dissipation, speed, and fluctuations. 
While there have been a number of studies on the relation between error probability, dissipation, and speed, 
somewhat less attention was paid to the fluctuations emanating from the dynamical process. 
Mallory \emph{et al.} concluded that fluctuations are the lowest priority in optimizing copy processes in comparison with the speed, cost, and accuracy \cite{Mallory2019}; however,
large fluctuations in the reaction cycles of transcription and translation contributing to higher variability in the protein copy number \cite{Soltani2016,DalCo2017} could be detrimental to the fitness of an organism \cite{Fraser2004,Hausser2019}.
Given that the DNA replication of developing animals is meticulously synchronized across the cells, disarray of cell cycle from other cells can be lethal \cite{Blythe2015,Djabrayan2019}.
As further discussed in the section \textbf{mRNA translation by \emph{E. coli} ribosome, 
the fluctuations of the copy process could display substantial variations depending on the proofreading dynamics of the copying enzyme (Fig.~\ref{fig:Q_copy}c).
}
Overall, fluctuations associated with biological copy processes also have to be suppressed down to biologically acceptable levels, and this aspect is naturally taken into account by calculating the uncertainty product $\mathcal{Q}$ of TUR.  
\\

{\bf DNA replication.} 
Of the three biological copy processes, 
the most precise one is DNA replication,
at the heart of which is the 
interaction between the polymerase and the exonuclease domains of the DNA polymerase (DNAP) \cite{Johnson1993}.
In a nutshell, the exonuclease domain of the DNAP executes the proofreading step by preferentially cleaving off incorrect nucleotides,
the incorporation of which slows down the action of the polymerase.
Through a Michaelis-Menten approximation of all the chemical reactions catalyzed by the DNA polymerase,
Gaspard has derived the dependence of the error probability, speed, and thermodynamic cost on the substrate concentrations \cite{Gaspard2016b,Gaspard2016a}.
Along similar lines, Igoshin, Kolomeisky and colleagues further explored the reaction dynamics of the DNAP, 
by explicitly modeling the switching of the DNAP between the polymerase and exonuclease states \cite{Banerjee2017,Mallory2019}. 
Results from both groups demonstrated that the copy errors were being suppressed at the expense of the speed and the thermodynamic cost. 
Furthermore, the analysis of rate constants of the wild type T7 DNA polymerase suggested that it was optimized for the speed rather than the accuracy. 

Evaluating the fluctuation and the uncertainty product $\mathcal{Q}$ of DNA polymerases together with the activity of exonuclease requires a careful analysis of the kinetics involved with the switching among the 
polymerizing, proofreading, and paused states of the polymerase \cite{Pineros2020,Hoekstra2017,Ibarra2009}.
However, since the precise knowledge of exonuclease mechanism still remain elusive, the analysis here is limited to the proofreading-free version of exonuclease-deficient T7 DNAP. 
The analysis of kinetic network of T7 DNAP finds that 
$\mathcal{Q}$ for T7 DNAP is suboptimal ($\mathcal{Q}\approx 10$) at [dNTP] $\approx 100$ $\mu$M, which is in the similar range of dNTP at the cellular condition $\mathcal{O}(10^2)$ $\mu$M $-$ $\mathcal{O}(1)$ mM \cite{Bochner1982,Buckstein2008,Schaaper2013}. 
Notably, the error probability is already saturated to its minimal value $\eta\approx 10^{-4}$ when [dNTP] $\gtrsim 10^{-11}$ M. 
This suggests that instead of the error probability, 
other dynamical properties of T7 DNAP can be further optimized.  
It is noteworthy that the dynamics of the T7 DNAP is sub-optimal in $\mathcal{Q}$ at its working condition.
\\

{\bf RNA polymerases. }
While synthesizing the RNA transcript complementary to the DNA sequence, 
the RNA polymerase (RNAP) translocates along the DNA sequence, maintaining a DNA bubble of 12-14 basepairs, and an 8-9 basepair RNA-DNA hybrid double strand \cite{abbondanzieri2005Nature,Chen10PNAS}. 
Similarly to the DNAP,
the polymerase and the exonuclease activities of the RNAP are combined together to suppress copy errors \cite{Sydow2009}. 
Upon incorporating incorrect nucleotides, the polymerase activity slows down and the RNAP converts to a backtracked state that can no longer incorporate a new nucleotide. 
Only after removal of the erroneous NTP, the RNAP is able to continue incorporating new nucleotides. 
Structural, biochemical, and theoretical studies have characterized both the mechanisms of nucleotide incorporation in the elongation complex and the proofreading, highlighting the strong sequence dependence on the error rate and the pausing frequency of the polymerase. 
The effect of the exonuclease-mediated proofreading reactions on the error probability have been quantitatively evaluated through kinetic modeling \cite{Mellenius2017}. 
Along with the fluctuations of the RNAP activity, analysis of RNAP dynamics in light of $\mathcal{Q}$ would also be of great interest. 
\\

{\bf mRNA translation by \textit{E.~coli} ribosome. }
The \textit{E. coli} ribosome is a well characterized system that employs KPM to suppress copy errors \cite{Blanchard2004,Rodnina2018}.
The ribosome synthesizes proteins from mRNAs by decoding the sequence information encoded in the `codons'.
For each of the 61 codons (excluding the 3 stop codons from the total of $4^3$ possible combinations), 
there exist potentially multiple `correct' tRNAs with the matching anti-codon, in complex with the encoded amino-acid (aa), elongation factor (EF), and GTP. 
The binding of the correct aa-tRNA-EF-GTP complex to the ribosome-mRNA complex initiates the reaction cycle through which the amino-acid is added to the elongating polypeptide sequence.
Similarly, aa-tRNA-EF-GTP complexes with incorrect amino-acids can undergo a parallel reaction cycle as the correct substrate, which leads to the incorporation of errors in the polypeptide sequence \cite{Rudorf2014}.
Free energy released from GTP hydrolysis is used to execute KPM (see Ref.\cite{Song2020JPCL} for the details of kinetic proofreading mechanism).   

The error probabilities associated with codon-anticodon pairings are already saturated to $\eta\approx 10^{-4} - 10^{-2}$ for all the codons with respect to the variations in [aa-tRNA] and [GTP] (see Fig.S6 in Ref. \cite{Song2020JPCL}). 
As shown in Fig.\ref{fig:Q_copy}b, the uncertainty products $\mathcal{Q}$ are again non-monotonic functions of both [aa-tRNA] and [GTP], and the corresponding cellular concentrations of [aa-tRNA] and [GTP] are found greater than the concentrations, [aa-tRNA]$^\ast$ and [GTP]$^\ast$, that give rise to the suboptimal values of $\mathcal{Q}$. 
Depending on the codon type, $\mathcal{Q}$ varies between 20 and 40 (Fig.\ref{fig:Q_copy}b). 
The mRNA translation is realized when ribosome translocates through a string of codons, accommodating correct types of aa-tRNA and forming new peptide bonds. 
The greater forward kinetic rates of the GTP hydrolysis and the polymerization along the correct cycle \cite{Wohlgemuth2011,Rudorf2014} lowers the error probability.

To study the process of mRNA translation of \emph{E. coli.} ribosome in a more realistic fashion, 
it is possible to consider an extended version of network model which translates 42 species of aa-tRNAs into 20 different amino acids. 
With information on the concentration of aa-tRNAs in the cellular milieu, 
Song \emph{et al.} classified them into cognate, near-cognate, and non-cognate types and simulated the \emph{E. coli.} ribosome-mediated translation of the \emph{tufB} mRNA sequence that encodes the 394 amino acid EF-Tu (Fig.~\ref{fig:Q_copy}c) \cite{Song2020JPCL}.
The effect of proofreading step on the ribosome dynamics as well as on the error probability  
can be assessed 
by modulating the polymerization current by multiplying a factor $\alpha_{\rm PR}$ to the associated rate constants (see the general kinetic network for proofreading depicted in Fig.\ref{fig:Q_copy}b).  
A similar perturbative analysis has previously been used to decipher which feature of the biomolecular process had been  optimized throughout evolution \cite{Banerjee2017,Mallory2019,Mallory2020a}.
Although rate constants are not easy to tune in experiments, 
such modification happens throughout the evolution by means of mutations to the ribosome, EF-Tu, and tRNA. 
Their simulation results demonstrated that the variations in the completion times of mRNA translation (the first passage times, $\mathcal{T}$) depends critically on the parameter $\alpha_{\rm PR}$ that modulates the polymerization current. 
For $\alpha_{\rm PR}=1$ corresponding to the wild type, the average speed of polymerization is found $\approx 16$ aa/sec and the error probability is $\eta\approx 10^{-3}$ (Fig.\ref{fig:Q_copy}c), in good agreement with those known from experimental measurements \cite{Young1976,Bouadloun1983}. 
Selecting the completion time $\mathcal{T}$ (first passage time) as the output observable, one can use a version of TUR derived for the first passage time \cite{gingrich2017PRL}:
\begin{align}
\mathcal{Q}=\Delta S_{\rm tot}(\mathcal{T})\frac{{\rm Var}(\mathcal{T})}{\langle \mathcal{T}\rangle^2}=\sigma_{\rm tot}\frac{{\rm Var}(\mathcal{T})}{\langle \mathcal{T}\rangle}\geq 2.
\end{align} 
While $\eta$ decreases monotonically with $\alpha_{\rm PR}$, 
$\mathcal{Q}$  is non-monotonic with $\alpha_{\rm PR}$ (Fig.\ref{fig:Q_copy}c), minimized near the wild type condition. 
At $\alpha_{\rm PR}=1$ and under the cellular concentrations of aa-tRNA and GTP, 
the uncertainty product for \emph{E. coli.} ribosome is $\mathcal{Q}= 45$ (Fig.\ref{fig:Q_copy}c). 
For the given kinetic parameters from the wild type (WT), $\mathcal{Q}$ is minimized to $\mathcal{Q}\sim 30$ at $\alpha_{\rm PR}\approx 5$. 
For $\alpha_{\rm PR}= 10^{-2}$, 
the translation times display a much broader distribution than that of the wild-type (Fig.\ref{fig:Q_copy}c). 
Thus, it could be argued that the extent of proofreading of the WT is in a proper range, that the fidelity of translation and the fluctuations of protein synthesis are simultaneously regulated. 
Fluctuations in the completion time for mRNA translation ($\langle\delta\mathcal{T}^2\rangle$) can be critical, as it is translated to significant variation in protein copy number.

For the \emph{E. coli.} ribosome, 
the kinetic rate constants of ribosomes have evolved to optimize the speed ($\langle J_{\rm pol}\rangle$) over the accuracy ($\eta$), while increasing the thermodynamic cost ($\mathcal{A}$) only slightly above the minimal cost 
necessary for the polymerization reaction. 
Importantly, similar conclusions were drawn by a number of studies, each of which evaluated the translation process of the \textit{E.~coli} ribosome using different kinetic reaction networks \cite{Wohlgemuth2011,Banerjee2017,Song2020JPCL}.
From the findings of maximized current, suppressed fluctuations, and moderate increase of $\mathcal{A}$ captured by the effectively sub-optimized value of $\mathcal{Q}$ $(\approx45-50)$ at $\alpha_{\rm PR}\approx 1$, while $\eta$ being determined at biologically acceptable levels  \cite{Pineros2020,Song2020JPCL}, 
it could be argued that the wild type ribosome operates near semi-optimal condition. 
\\

\begin{figure}[t]
\begin{center}
	\includegraphics[width=.48 \textwidth]{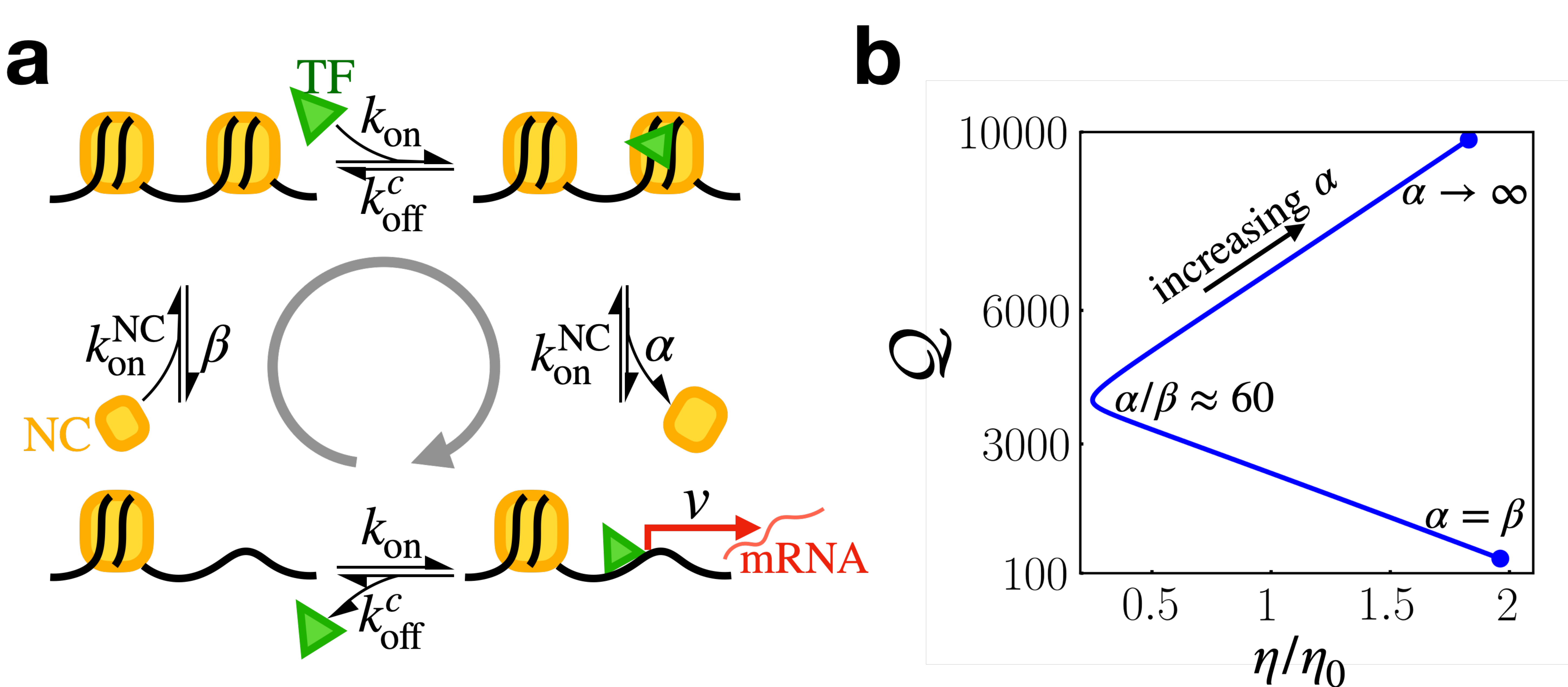}
	\caption{$\mathcal{Q}$ and transcription noise in non-equilibrium gene regulation by transcription factors (TFs). 
	(a) Schematic of a gene regulation by a combination of the nucleosome (NC) and TF occupancy. 
	Transcriptional activity is assumed to occur only when the regulatory sequence is released from 
	the NC and bound to the transcription factor (TF) \cite{Shelansky2020}. 
	A parallel network for incorrect gene expression also exists, in which the TF unbind with rate $k^i_{\rm off}$.
	(b) The error probability ($\eta$) and $\mathcal{Q}$ of correct gene expression plotted as functions of $\alpha$.
	When the NC is removed preferentially in the TF bound state (i.e. $\alpha>\beta$), 
	the error in gene expression can be reduced below $\eta_0 \equiv [1+k^i_{\rm off}/k^c_{\rm off}]^{-1}$.
	Other than $\alpha$, the remaining parameters are set as follows: 
	$k_{\rm on} = 1~{\rm sec^{-1}}$,
	$k^c_{\rm off} = 1~{\rm sec^{-1}}$,
	$k^i_{\rm off} = 100~{\rm sec^{-1}}$,
	$k^{\rm NC}_{\rm on} = .2~{\rm sec^{-1}}$,
	$\beta = 0.01~{\rm sec^{-1}}$, and
	$v = 0.01~{\rm sec^{-1}}$.
	}
	\label{fig:nucleosome}
	\end{center}
\end{figure}

\subsection*{Gene regulation by transcription factors}
Gene expression is regulated in large part by the binding of transcription factors (TFs) to the regulatory regions of DNA. 
The specificity by which TFs can activate target genes 
is originated from the discriminatory binding of TFs to the non-specific regulatory regions of the DNA.
With an assumption that the relative binding affinities of the TF to the target and the non-target sequences differ only by their respective dissociation rates, $k^{c}_{\rm off}$ and $k^{i}_{\rm off}$,
the minimal error probability of transcription can be approximated by $\eta \approx \eta_0 = [{1+k^{i}_{\rm off}/k^{c}_{\rm off}}]^{-1}$. 
Shelansky and Boeger recently proposed 
a mechanism in gene regulation by TFs and nucleosomes that can reduce $\eta$ at the expense of extra free energy cost \cite{Shelansky2020}.
Nucleosomes, the regulatory structures that bind and unbind reversibly from DNA, are generally known to suppress transcriptional activity in the bound state.
In Shelansky and Boeger's model, transcription is assumed to occur only when the DNA is bound to the transcription factor, and free from the nucleosome (Fig.~\ref{fig:nucleosome}a).
Briefly, when the nucleosome can be more easily removed from the TF-bound regulatory sequences than from the TF-absent ones ($\alpha>\beta$ in Fig.~\ref{fig:nucleosome}a),
a kinetic proofreading-like mechanism can reduce the minimum error rate below $\eta_0$.
The nucleosome removal in the TF-bound sequences is predicted to be driven by the activity of ATP-consuming chromatin remodelers, 
which are recruited by the TF \cite{Zhou2016b,Shelansky2020}.

Fig.~\ref{fig:nucleosome} plots the error probability and $\mathcal{Q}$ with an increasing thermodynamic drive. 
In parallel to Eq.\ref{eq:error_prob}, the error probability is defined as $\eta\equiv J^i_{\rm trans}/(J^i_{\rm trans}+J^c_{\rm trans})$, where $J^i_{\rm trans}$ and $J^c_{\rm trans}$ are the reaction currents associated with the incorrect and correct gene expression. 
Next, the noise in the correct gene expression is defined by the Fano factor $\lambda \equiv \langle (\delta J^c_{\rm trans})^2 \rangle / \langle  J^c_{\rm trans} \rangle$. 
To evaluate the uncertainty product $\mathcal{Q}$, we assume that the nucleosome removal step ($\alpha$) entails the unwinding of $\sim$20 basepairs of DNA, with every 2 basepairs requiring 1 ATP based on single molecule studies \cite{Shundrovsky2006, Zhang2006, Sirinakis2011, Deindl2013}. We additionally assume that the transcription initiation step ($v$) requires $\sim$20 ATPs in order to melt the DNA into the open complex \cite{Lynch2015a}. Assuming that the free energy of ATP hydrolysis is $\sim$20 $k_BT$, we can compute the semi-empirical free energy cost associated with transcription, $\mathcal{A}$, by accounting for all the ATP hydrolysis reactions occurring at the correct and incorrect DNA binding sites. 

To increase the thermodynamic drive, we increase the nucleosome unbinding rate from TF-bound sequences ($\alpha$), 
while keeping the rest of the parameters constant.
$\mathcal{Q}$ increases monotonically with $\alpha$, 
while the error probability ($\eta$) is minimized at around $\alpha/\beta\approx 60$.
Further analysis in the framework of the TUR will be useful in quantifying trade-off relations among the error probability, transcription noise, and the thermodynamic  cost of gene regulation by TF binding.

\begin{figure}[t]
\begin{center}
	\includegraphics[width=1.0\linewidth]{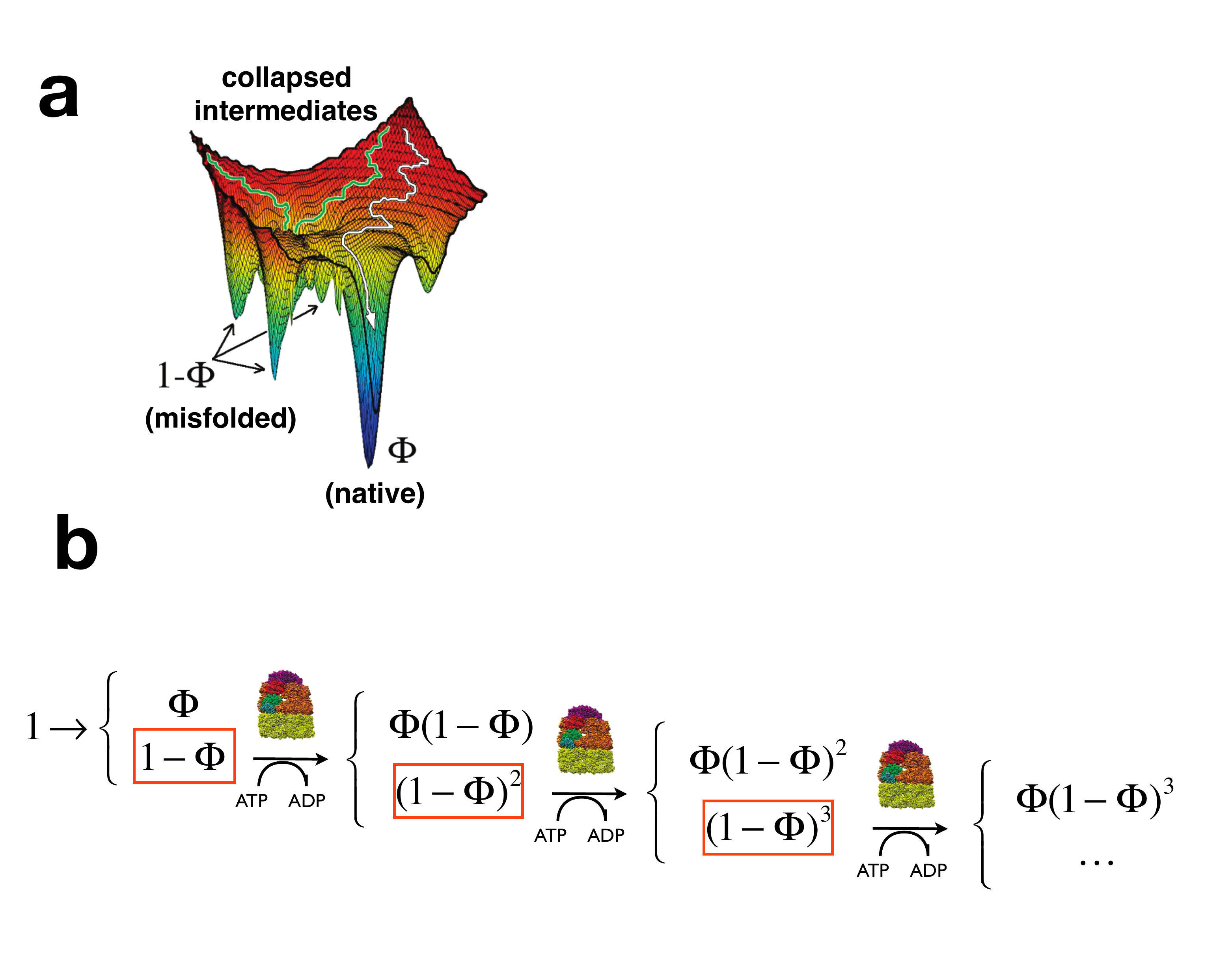}
	\caption{Molecular chaperone as an error-reducing machine. 
	{\bf (a)} Rugged folding landscape of biomolecules that visualizes the native and misfolded basins of attraction. 
	As a result of spontaneous folding, molecules are partitioned with the fraction of $\Phi$ and $1-\Phi$ to native and misfolded states, respectively. 
	{\bf (b)} Iterative annealing by GroEL.  
	The figure (a) was adapted from Ref.\cite{Thirum05Biochem}. 
}
	\label{fig:chaperone}
	\end{center}
\end{figure}

\subsection*{Molecular chaperone-assisted folding of proteins}
In an appropriate environmental condition, small single domain proteins in general can reversibly fold and unfold, and reach their native state within biologically relevant time scales \cite{Anfinsen75APC}. 
Yet, there are still a class of proteins that are prone to misfold and aggregate, whose presence can be detrimental to the organisms. 
For such proteins (e.g., Rubisco, malate dehydrogenase \cite{Thirumalai01ARBB,chakrabarti2017PNAS}), only a small fraction ($\Phi\ll 1$, $\Phi\approx 0.05$ for Rubisco \cite{Todd96PNAS}) of the population can reach their native state, and the remaining fraction of population ($1-\Phi$) are kinetically trapped in misfolded states (Fig.\ref{fig:chaperone}). 
Molecular chaperones \cite{Thirumalai01ARBB,Thirum05Biochem,chakrabarti2017PNAS,Mugnai2020RMP,Goloubinoff18NCB}, which employ the free energy sources ubiquitous in cells to change their conformations and interact with the molecules in misfolded states, can change the population entirely and sustain the cellular environment in good condition.    

One of the most well-studied protein chaperones, bacterial GroES-GroEL chaperonin system interacts exclusively with the misfolded population of proteins, providing them with another chance to repeat the folding process. 
As a result of an initial interaction of the chaperone with misfolded protein population, 
out of the misfolded population $1-\Phi$ from the first round of folding process, $\Phi(1-\Phi)$ would fold into the native state, and $(1-\Phi)^2$ would be again trapped in the misfolded states. 
When this process is repeated $N(=t/\tau_0)$ times where $\tau_0$ is the time associated with a single cycle and $t$ is the time duration, 
the fraction $(1-\Phi)^N$ still remains misfolded and hence the $1-(1-\Phi)^N$ are folded. 
The fraction of native population increases from its originally small yield $Y_1=\Phi(\ll 1)$ at $N=1$ to 
\begin{align}
Y_N=1-(1-\Phi)^N\xrightarrow{\Phi\ll 1} 1-e^{-\Phi t/\tau_0}. 
\end{align} 
Via this mechanism, called the iterative annealing mechanism (IAM) \cite{Todd96PNAS,Thirumalai01ARBB,Hyeon2013JCP,chakrabarti2017PNAS,korobko2020Elife}, 
the native yield of unity can be finally reached when $N\rightarrow \infty$ or $t\rightarrow\infty$.  
From the perspective of information processing, molecular chaperones are another elegantly designed error reducing machinery. 
For every cycle, which lasts about 2 sec \cite{Ye13PNAS,Fei14PNAS}, 
GroEL made of two heptameric rings presumably consumes at least $3-4$ ATPs for each ring, which amounts to $\approx$ $60-80$ $k_BT$ of dissipation.   
Furthermore, since successful conversion of misfolded to folded state is not guaranteed at each cycle, the dissipation per cycle estimated above is only a lower bound of the estimate. 
In fact, $\tau=\tau_0/\Phi$ is the full conversion time to the native state. Given that $\tau_0\approx 2$ sec and $\Phi=0.05$ for Rubisco, $\tau=40$ sec. 
Since the number of successful conversions to native state usually obeys Poisson statistics, the Fano factor of the net fraction of conversion $\Delta Y_N(t)$ is $\lambda=\langle(\delta \Delta Y_N)^2(t)\rangle/\langle \Delta Y(t)\rangle\approx \mathcal{O}(1)$. 
Thus, a rough estimate of $\mathcal{Q}$ for GroEL assisted protein folding is $\mathcal{Q}\gtrsim  \tau/\tau_0\times (60-80)\approx 2\times 10^3$. 

\begin{figure}[t]
\begin{center}
	\includegraphics[width=1\linewidth]{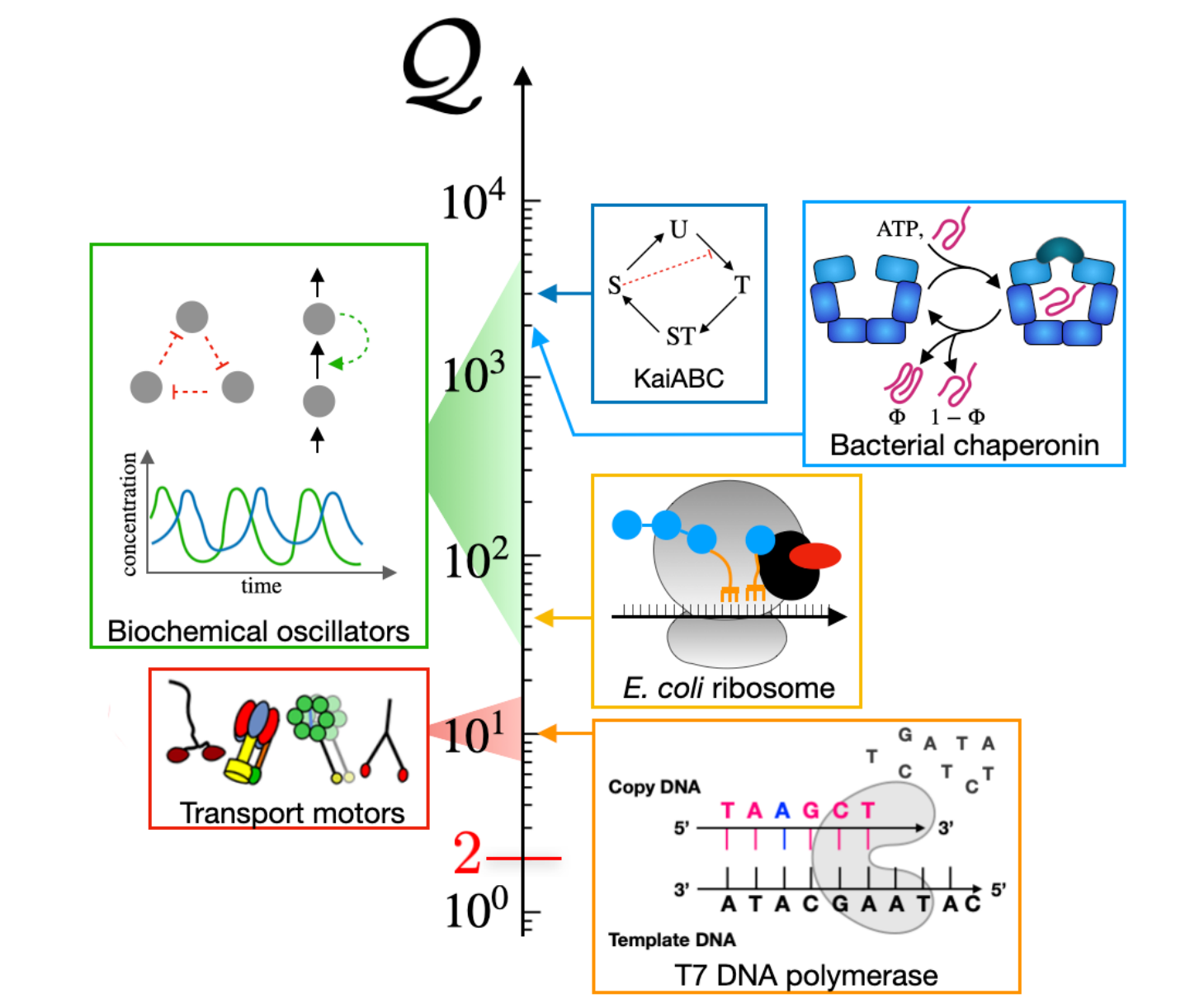}
	\caption{A spectrum of uncertainty product ($\mathcal{Q}$) computed for transport motors \cite{Hwang2018JPCL}, T7 DNA polymerase \cite{Song2020JPCL,Pineros2020},
	\textit{E. coli} ribosome \cite{Song2020JPCL,Pineros2020}, molecular chaperone, and biochemical oscillators \cite{Marsland2019}. The theoretical lower bound of TUR is specified at $\mathcal{Q}=2$. 
	The range of $\mathcal{Q}$ for the family of molecular motors and biochemical oscillators are shaded in
	red and green, respectively \cite{Hwang2018JPCL,Marsland2019}.
	}
	\label{fig:Fig5}
	\end{center}
\end{figure}

\section*{Conclusions}
Under evolutionary pressure, biological processes are often confronted with situations in which to balance between a number of competing options. 
Here, we have reviewed the recent studies on the trade-off relations of these features in biological motors and biological dynamics involved with information processing, from the perspective of TUR which offers the uncertainty product $\mathcal{Q}$ as a measure of the optimality integrating the cost and precision of the processes.
Importantly, the physical lower bound of $\mathcal{Q}$ provides an absolute scale onto which we can map the efficiencies of diverse biomolecular processes.
Biological motors, specialized for cargo transport, are found to operate close to the lower bound ($\mathcal{Q}=2$) even at NESS, with $\mathcal{Q}\approx 7-15$ \cite{Hwang2018JPCL}. 
Compared with biological motors that utilize ATP hydrolysis free energy ($\sim 20$ $k_BT$ $\approx $ 0.8 eV), synthetic nano-motors \cite{kudernac2011Nature} that use $>3$ eV UV-light source as the driving force are expected to have at least several-fold greater $\mathcal{Q}$. 
The exonuclease-deficient T7 DNA polymerase operate at $\mathcal{Q}\approx 10$, and the ribosome operates at $\mathcal{Q}=45-50$ \cite{Pineros2020,Song2020JPCL}. 
The value of $\mathcal{Q}$ for molecular chaperones is estimated to be rather large, $\mathcal{Q}\gtrsim \mathcal{O}(10^3)$, mainly due to the large cost of operating the reaction cycle of chaperones. 
Although it was not discussed in this article, 
Marsland \emph{et al.} have evaluated the values of $\mathcal{Q}$ for several biochemical oscillators by selecting the oscillation period as their output observable of interest. 
Some of the biochemical oscillators severely underperform the TUR bound; for example, the KaiABC system works at $\mathcal{Q} \gtrsim O(10^3)$ \cite{Marsland2019}. 
For biological motors, both thermodynamic cost and precision of the processes are valuable quantities to be balanced, giving rise to relatively small value of $\mathcal{Q}$ 
near the physical lower bound of $2$. 
In contrast, for the cases of molecular chaperones and biochemical oscillators, the precision of the processes is in the highest priority, at the expense of large thermodynamic cost. 
Along with the spectrum of $\mathcal{Q}$ recapitulating our survey on various processes in this article (Fig.~\ref{fig:Fig5}), the $\mathcal{Q}$ values of more number of other dynamical processes will be of help to better glean the design principles underlying life sustaining cellular processes.

\acknowledgements{This work was supported by the KIAS individual Grants CG067102 (YS) and CG035003 (CH) at Korea Institute for Advanced Study.
We thank the Center for Advanced Computation in KIAS for providing computing resources.}

\section*{Data Availability}
The data that supports the findings of this study are available within the article. 

\bibliographystyle{jcp}

\end{document}